\documentclass[preprint2,longabstract]{aastex62}

\usepackage{enumerate}
\usepackage{color}

\def\grad{$^\circ$}
\def\it{\sl}
\def\degs{\ifmmode ^{\circ}\else$^{\circ}$\fi}
\def\amin{\ifmmode ^{\prime}\else$^{\prime}$\fi}
\def\asec{\ifmmode ^{\prime\prime}\else$^{\prime\prime}$\fi}

\def\degs{\ifmmode ^{\circ}\else$^{\circ}$\fi}
\def\amin{\ifmmode ^{\prime}\else$^{\prime}$\fi}

\def\eqalign#1{\null\,\vcenter{\openup1\jot \m@th
   \ialign{\strut\hfil$\displaystyle{##}$&$\displaystyle{{}##}$\hfil
   \crcr#1\crcr}}\,}
\usepackage{color}
\newcommand{\gt}[1]{{\color{blue} \textbf\small{#1}}}

\newcommand{\sgr}{V1082\,Sgr}
\def\apgt{\ {\raise-.5ex\hbox{$\buildrel>\over\sim$}}\ }
\def\aplt{\ {\raise-.5ex\hbox{$\buildrel<\over\sim$}}\ }

\shorttitle{Magnetic  pre-CV V1082 Sgr}
\shortauthors{Tovmassian et al.}

\begin{document}

\title{$K2$ study of the Magnetic Precataclysmic Variable V1082\,Sagittarius}

\author{Gagik Tovmassian}
\affiliation{Instituto de Astronom\'{\i}a, Universidad Nacional Aut\'onoma
de M\'exico, Apartado Postal 877, Ensenada, Baja California, 22800 M\'exico}
\email{gag@astro.unam.mx}
\author{Paula Szkody}
\affiliation{Department of Astronomy, University of Washington, Box 351580, Seattle, WA 98195, USA}
\author{Ricardo  Yarza}
\affiliation{Instituto Tecnol\'ogico y de Estudios Superiores de Monterrey,  64849 Monterrey, N.L., M\'exico}
\author{Mark Kennedy}
\affiliation{Jodrell Bank Centre for Astrophysics, School of Physics and Astronomy, The University of Manchester, Manchester M13 9P
}

\begin{abstract}
We present a long-term light curve of the precataclysmic variable (CV) V1082\,Sgr obtained by the $K2$-mission over the course of 81 days.
We analyze the entire complex light curve as well as explore several sections in detail with a sliding periodogram.
The long dataset allows the first detection of the orbital period in the light curve, as well as the confirmation of cyclical variability on a 
longer timescale of about a month.  A portion of the light curve in deep minimum reveals 
a clean, near-sinusoidal variability attributed to the  rotation of   the spotted surface of the donor star. We model that portion of the light 
curve assuming that the donor star grossly under-fills its Roche lobe,  has cool spots similar to a chromospherically active, 
slightly evolved early K-star, and  might be  irradiated by the X-ray beam from the magnetically accreting white dwarf.
The fast variability of the object in the active phases resembles the light curves of magnetic CVs (polars). 
\end{abstract}


\section{Introduction}

Cataclysmic Variables  (CVs) are interacting close  binary systems consisting of a red star filling its corresponding Roche lobe 
and losing matter to a white dwarf (WD) companion \citep{1995CASSS...5.....H}.  In systems with a strongly magnetic WD 
the matter falling into the potential well of the WD is captured and channeled onto  the magnetic pole(s).  There are also a  handful of 
binaries  known as prepolars, which are considered as part of the CV family, as there is  evidence of  accretion onto the 
WD in these systems \citep{2005ASPC..330..137W,2005ApJ...630.1037S}.  All of them contain late M dwarfs, and the orbital periods of the systems 
are comparable  to typical orbital periods for CVs (i.e.  less than 6 hr). 

However, where the prepolars differ dramatically from CVs is in their accretion rates. Prepolars have been observed to accrete at a rate 
that is two orders of magnitude less than the typical rate in CVs. In fact, prepolars are technically not CVs, 
since the accretion geometry in CVs typically requires the secondary to be overflowing its Roche lobe, while 
prepolars are detached binaries. Instead, accretion in prepolars is assumed to arise from a  coupling of the 
magnetic fields of both stars in the binary that draws matter lost by the wind from the donor star to accrete onto the magnetic pole of the WD
\citep{2012ApJ...758..123W,2015SSRv..191..111F}.  
There should be counterparts of prepolars with donors of earlier spectral types.  So far, only two or three 
candidates have been proposed and the jury is still out as to whether these systems are truly prepolars  
\citep{1988ApJ...331L..29R,2017ASPC..509..489T}. 

The $K2$ mission commenced after the original {\sl Kepler} mission was terminated for technical reasons \citep{2014PASP..126..398H}. 
It provides precision photometry  of fields concentrated around the ecliptic plane. $K2$ provides data at long 30\,m and short 1\,m cadences over 
$\sim$80 day intervals with 
precision close to that of the original {\sl Kepler} mission.   We  proposed observations of   \sgr, an object that demonstrates a large amplitude variability 
on very different time scales \citep{2016ApJ...819...75T}. Previous long-term photometry revealed that the system is active, or bright most of the 
time with occasional deep minima. The variability appears to be quasiperiodic with cycles roughly 29 days long. No definite periodicities were 
found at that or shorter periods. The orbital period of 20.82\,hr was determined from the spectroscopic analysis of radial velocity (RV)
variations \citep{2010PASP..122.1285T}.  
The orbital period is  well above the 0.4\,day upper limit at which a zero age main sequence red dwarf would be filling its corresponding Roche lobe in a 
binary with an average mass WD primary.  The donor  star is clearly visible at $ 50-100\%$ of the total flux around 5000\AA\ depending 
on the accretion activity state.  

Although the object 
often shows strong emission lines resembling a CV,  only a spectrum of a K\,2 star is visible in the optical range during the times of minima
with a lack of additional radiation in the continuum or spectral lines. This further supports the idea that the companion is underfilling its   Roche lobe,
as these minima are difficult to explain with a Roche lobe overfilling scenario. 
However, the precise determination of spectral and particularly the
luminosity class of the donor star based on the low-resolution and highly variable spectra was inconclusive. 

The secondary is chromosperically active as can be inferred from the presence of narrow H$_\alpha$ and \ion{Ca}{2} emission lines 
\citep{2016ApJ...819...75T} 
when the accretion is halted.   When the accretion turns on, the emerging strong emission lines are symmetric and of low RV
suggesting that they are not formed in an accretion stream characteristic to ordinary polars. 
Hence,  \citet{2016ApJ...819...75T,2017ASPC..509..489T} suggest that V1082\,Sgr is a prime candidate for one of  the hard sought detached 
binaries with a magnetic WD and a magnetically active K-star. 

V1082 Sgr has also been studied at X-ray wavelengths by \citet{2013MNRAS.435.2822B} who showed that \sgr\ is  highly variable  in X-rays,  
with variations on a wide range of time scales from hours to months. The length of the observation was not sufficient to cover the unusually long orbital period of 
the system.  However,  \citet{2016ApJ...819...75T} showed that the observed X-ray flare may be related to a changing viewing angle
of the accretion column over the orbital period during one of the system's accretion driven stages. 
A complex  fit to the X-ray spectrum reinforces the proposition  that  the plasma reaches typical temperatures achieved in a
magnetically confined accretion flow, where a standing shock is
formed at the poles of a compact star.

Here we present the  analysis  of $K2$  observations  of \sgr, which provides new information on the accretion process  
in this system and adds to the  evidence for the magnetic nature of the binary.  Some parallels with the X-ray results can be drawn.

\clearpage

\section[]{Observations}

\begin{figure}[t]
\centering
\includegraphics[width=8.5cm,  clip]{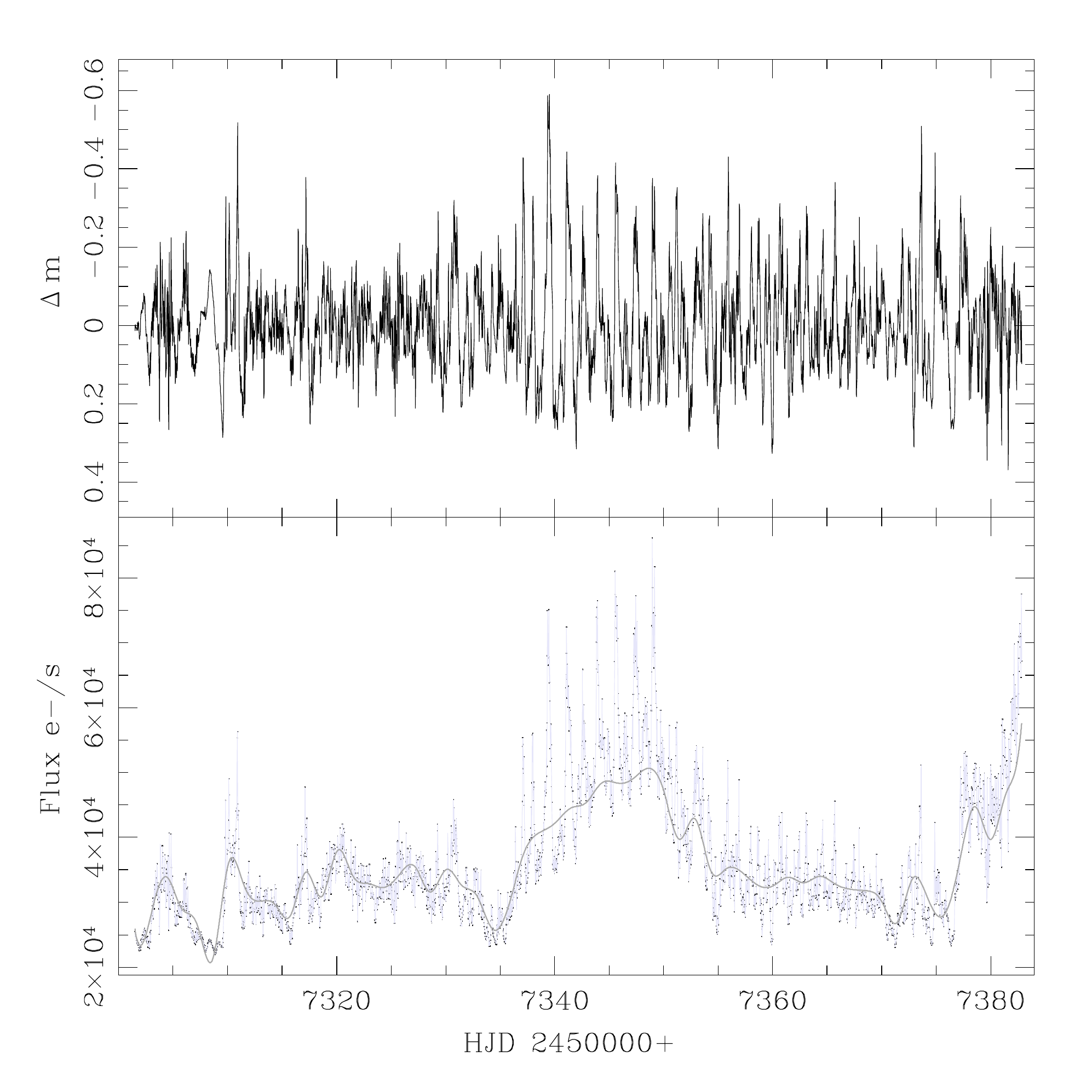}
  \caption{ Light curve of \sgr. Bottom panel: the individual measurements are plotted as tiny dots, connected by thin line.  The thick line is a cubic polynomial  
  fit  to the data. Top panel: the residual light curve after subtraction of the fit and converted to a logarithmic scale.  }
  \label{fig:lc}
\end{figure}

\sgr\ was observed by the {\sl Kepler} spacecraft as part of Campaign 7 of the $K2$ mission from 2015 October 4 
to December 26 (BJD =245\,57301.4 -- 57382.8; MJD 2467.2 -- 2549.8 ).
The integration time of the {\sl Kepler} spacecraft is 6.02\,s. Since it is unfeasible to save every single data point taken of an object with this short an integration 
time (due to the limitations of bandwidth when downloading data from the spacecraft), {\sl Kepler} data are averaged over two different time spans -- 
short cadence and long cadence. For short cadence data, exposures are averaged over approximately 1 minute, while long cadence data correspond 
to 30 minutes of data. As such, the short cadence data can be treated as a subset of long cadence data, since they are both derived from the same set of 6\,s exposures.

We obtained  both  long  and short cadence data (exposures of
29.4\,m and 58.8\,s respectively)  for \sgr. 
The data are in the form of integrated  photoelectrons collected during either a one or  30\,m observation. Each data point in the time series is the direct sum 
of  counts within a predefined aperture. The apertures are constructed to maximize the signal-to-noise ratio of the light curves and take into account the 
varying pixel response function across the focal plane \citep{2010ApJ...713L..97B}.  

The light curves obtained by the in-house pipeline reduction  were downloaded from the Mikulski Archive for Space Telescopes.  
For the short cadence light curve we acquired data processed by the Everest  pipeline \citep{2016AJ....152..100L} which aims to increase the precision of the photometry
designed for  exoplanet detection. However we found that the procedure misinterpreted the overall variability of the object and treated them as 
undesirable trends, removing features which we believe are part of the intrinsic variability of this peculiar binary. Hence, we use in the following analysis the raw fluxes
corresponding to the {\scshape{sap\_\,flux}} in the long cadence dataset.  In general, the variability demonstrated by this object far exceeds the stringent  
precision limits of the {\sl Kepler/K2} mission, designed to detect microvariability produced by transient extra-solar planets.  
The initial 1.2 days of data,  as well as outliers found upon visual inspection as not belonging to the 
intrinsic variability of the object, were removed.
 
 \begin{figure}[t]
 \includegraphics[width=8cm, bb=20 160 580 700, clip]{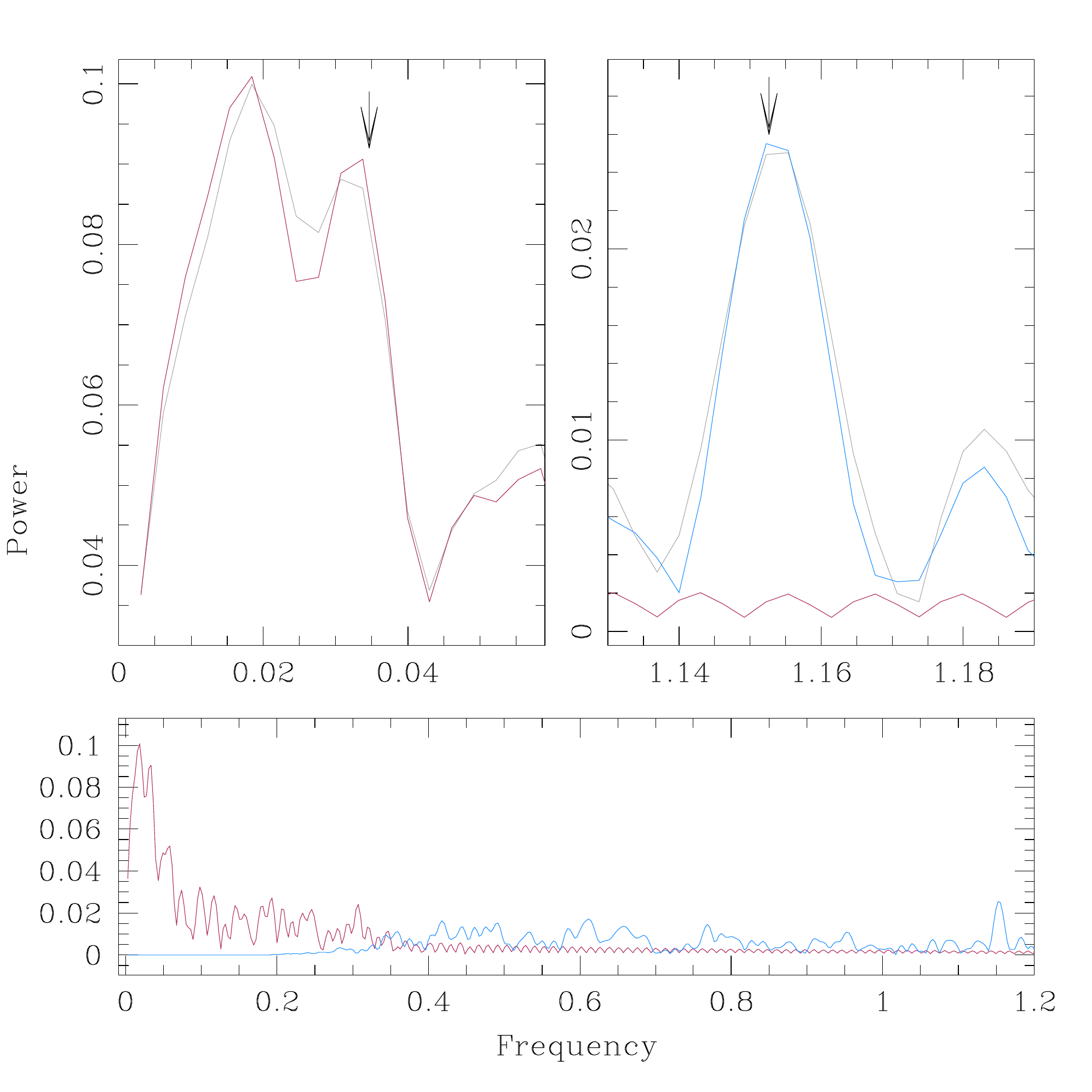}
  \caption{ Power spectra of the polynomial fit to the light curve (red curve) and to the residual light curve (blue curve) are presented in the bottom panel.  
  The zoomed-in portions of the power spectra with prominent peaks are presented in upper panels.  The arrow in the upper left panel indicates 29\,days 
  cyclical variability of the object. The arrow in the upper right panel corresponds to the spectroscopic orbital period of the binary. The gray curve on both of the top panels
  represents a power spectrum of the original observed light curve.   The frequency is in days$^{-1}$.
 }
  \label{fig:pws}
\end{figure}

Additional 40 day monitoring of the object was obtained with the  1.5\,m telescope of Observatorio Astron\'omico Nacional at San Pedro M\'artir (OAN SPM) 
equipped with the RATIR instrument 
\citep{2012SPIE.8446E..10B,2012SPIE.8444E..5LW} and operating robotically.  The data were acquired in 
Bessel-V and near-IR  $J$ and $H$ bands  and differential photometry was used to obtain the final magnitudes. An automatic pipeline procedure 
implemented in python was used to perform preliminary tasks of bias 
subtraction, flat-fielding, and cosmic-ray removal. The  pipeline also conducts astrometric calibration and sky subtraction for infrared (IR) images. 
Aperture photometry  was done using IRAF {\sl aphot} package \citep{1986SPIE..627..733T}. 
For more detailed information on these and simultaneous spectroscopic 
observations see \citet[][hereafter Paper\,II]{paper2}

\section{ Long cadence data}

In the bottom panel of Figure\,\ref{fig:lc} we present the $K2$ long cadence light curve of the object. Individual measurements are plotted as tiny dots that are connected to 
illustrate the rapid, large-scale variability. The light curve shows two intervals of increased brightness, with the second one cut short by the termination of the observation.  
In addition to this long-term variability,  short-term variability is evident. The amplitude of short-term variability seems to be  higher when the overall brightness is higher 
(8-10\% versus 12-18\%).   We fitted the light curve with a high-order polynomial (cubic spline), shown as a thick curve in the same panel, 
in order to separate long-term and short-term variability. 
The high-amplitude and high-frequency peaks  
were excluded from the fit to study the underlying gross variability. The  polynomial order was selected to eliminate all variability over 0.33\,c/day frequency 
(corresponding to 3 days periodicity).   The result of the subtraction of the fit from the original data is shown in the top panel of Figure\,\ref{fig:lc} converted to 
a logarithmic scale.

 \begin{figure*}[!t]
\centering
 \includegraphics[width=18cm, bb=20 140 580 700, clip]{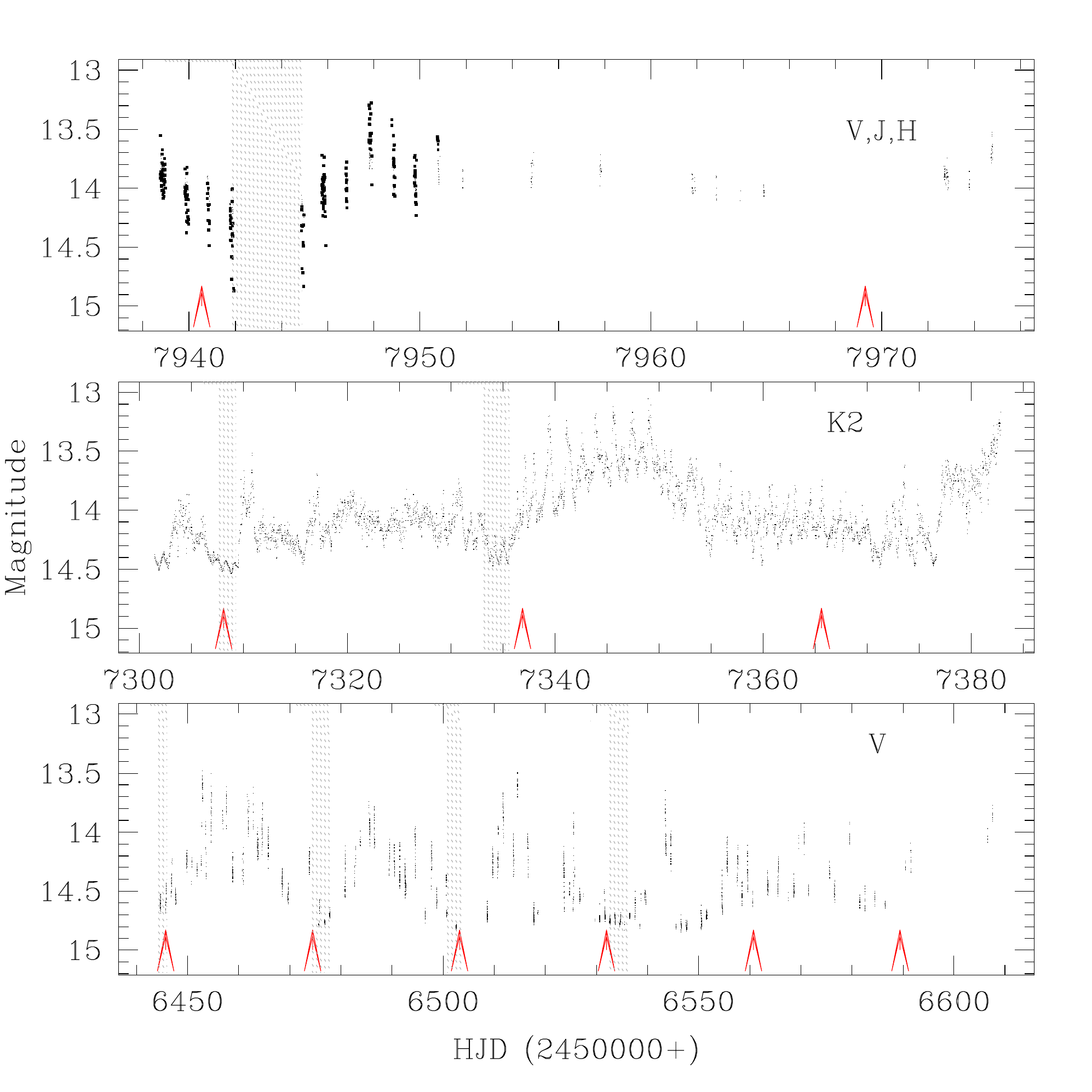}
  \caption{All available photometric observations of \sgr\ in different filters. 
  The optical $V$-band light curve from \citet{2016ApJ...819...75T} is presented in the bottom panel,  
  the K2 light curve in the middle  panel, and 
  the new photometric data in $V$ and near-IR $J-$ and $H$-bands are plotted in the top panel. 
  The magnitudes in $y$-axes reflect $V$ magnitude and bolder points correspond to the $V$-band. 
   Light curves in the other two bands are shifted roughly to visually match the former. 
   They are presented to cover a longer time interval and to demonstrate the variable pattern.  They are plotted in the same 
   magnitude scale, but the amplitude of variability in IR is smaller, hence they do not precisely overlap.\\
  Red arrows indicate moments of minima $\sim29$ day cyclical variability calculated from the arbitrary first occurrence. The shaded
   areas in the bottom and two upper panels indicate actual minima detections.  }
  \label{fig:xlc}
\end{figure*}

The fit and the residuals were subjected to  period analysis along with the unaltered light curve provided by the $K2$ mission. Taking into  account  
the even time distribution of the considered data, a simple discrete Fourier transformation method was used to calculate power spectra \citep{2005CoAst.146...53L}.
The power spectra are presented in Figure \,\ref{fig:pws}. In the bottom panel a range of frequencies is presented in which significant 
peaks are detected. The red curve is the power spectrum produced by the fit to the data and the blue curve corresponds to the residual light curve 
containing high-frequency variability.  The easily recognizable peaks in the power spectra are the $f_c\approx0.035$\, c/days and  
$f_c=1.153$\, c/days corresponding to  cyclical variability with roughly 29 days recurrence time  and  orbital period 0f 20.8\,hr, respectively. 
These peaks are shown in the insets of Figure \,\ref{fig:pws}. The black line in the upper panels is the power spectrum calculated with the 
unaltered light curve, demonstrating that we did not lose any information in the process of separating 
the low- and high-frequency variability. The orbital period was not detected  in the previous ground-based photometry  of \citet{2016ApJ...819...75T}. 
The low-frequency peak corresponding to the long-term brightness cycles  of the object is rather strong, although it is based only on 2.8 cycles enclosed 
within  the duration of observations. The strongest peak in the power spectra corresponding to $f=0.0184$\, c/days (54 days) is probably 
an artifact caused by the  brightening of the object around HJD\,2457350. 

Since  the duration of the $K2$ observation is certainly not enough to prove the continuity of the 29 day cycles, we combined all of the available photometry of the object in one plot.  
The composite light curve is shown in the three panels of Figure\,\ref{fig:xlc}, set in chronological order from bottom to top. 
In the bottom panel the observations in $V$-band reported in  \citet{2016ApJ...819...75T} are plotted, 
The second panel presents $K2$ data converted to $V$ magnitudes.  
The fluxes were converted into magnitudes according to guidelines of the $K2$ mission \citep{2016AJ....152....5D}. 
We also use the conversion of $K2$ magnitudes to $V$ as prescribed by \url{http://keplerscience.arc.nasa.gov} for a K0 - K5 star 
by adding a maximum $Kp -V=-0.29$ mag color correction.
Bessel-V and IR $J-$ and $H$-band light curves  from RATIR are plotted in the top panel.  The $y$-axes of the top panel correspond to 
magnitudes in the $V$-band with the data plotted as  thick points. The  $J$ and $H$ data are over-plotted as tiny points; 
they  were shifted to concur with the $V$-band  light curve purposefully to complement it for a longer time interval and underline the general trend.

The full amplitude of variability detected in the previous ground-based observations reaches $\approx1.5$\,mag,  
changing from 13.6 at the maximum 
to 14.8 mag at the minimum in the $V$-band (bottom panel).  A similar scale of variability is observed in the $K2$ data, but the variability band is brighter by about 0.3 mag, 
even after the maximum color correction by 0.29 mag (second from the top).  The disparity is a little large for comfort, but apparently 
there is no direct manner to relate measurements in the much narrower Johnson filters compared to the $K2$ bandpass.

The composite light curve covers over 1000 days. When subjected to a period analysis after normalization of each band to its faintest 
magnitude, a peak at  $f_c=0.0348$\, c/days frequency corresponding to 28.8 days is the strongest. The red arrows 
The moments of so called "deep minimum" are marked as shaded regions in the plots. This long-term variability is not strictly periodic, and that the
brightness of the object does not always reach a deep minimum. The deep minima are short-lived compared to the length of the cycle and their duration is variable
(please note that the horizontal scales ($x$-axes) of the panels are significantly different).

\begin{figure*}[t]
 \includegraphics[width=12cm, bb=70 40 580 720, angle=-90, clip]{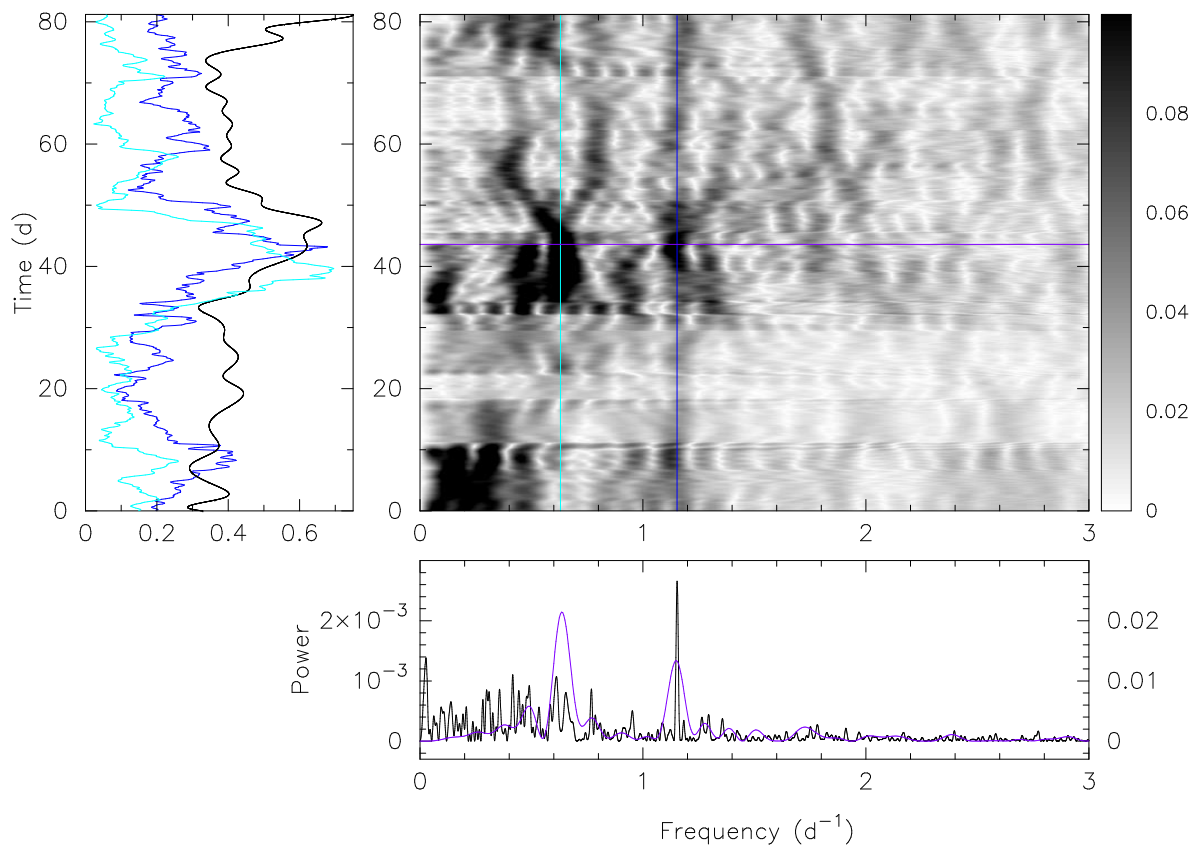}
  \caption{ Low-frequency end of the wavelet is presented in the central panel (see text for the description). In the bottom panel the power 
  spectra of the $K2$ light curve is presented with a black line. The dark violet line represents the power spectrum corresponding to the cut in the 
  wavelet indicated by the horizontal line of the same color. In the left panel the light curve of the object is plotted with a black line, and the dark 
  and light blue curves correspond to the intensity of frequencies marked by the vertical lines of corresponding colors. The dark blue line
  corresponds to the orbital frequency, while the light blue is the strongest, but transitional, peak at 0.63 cycles per day frequency.   
 }
  \label{fig:wvlt1}
\end{figure*}

Additional coverage from  the SPM falls in line with the rest of the accumulated data on the cyclical variability by providing  near-IR data. 
It is not dense enough to explore the orbital variability.   
The $V-J$ and $J-H$ color calculated around HJD\,2457942 from nearly simultaneous 
exposures are $1.85\pm0.05$  and $0.55\pm0.04$,  respectively. 
These values correspond to a K\,3-K\,4\,V star
 \citep{2013ApJS..208....9P},
which is cooler than K\,2 determined  from the spectral classification \citep{2016ApJ...819...75T}, but  once again 
confirming that in the deep minima  the 
donor star is the main contributor of light.  A possible explanation for the spectral class difference may lie with the IR-excess 
related to the presence of cyclotron 
radiation often observed in polars \citep{2015ApJS..219...32H}, which is beyond the scope of this paper.
In the maximum brightness around HJD\,2457948 
the increase is mostly in the optical light with  $V-J$ decreasing to $1.54\pm0.02$ while $J-H$ stays nearly constant.
For more on IR magnitudes and color index variability please see Paper\,II.

\section{Short cadence data}

\begin{figure}[!t]
\centering
 \includegraphics[width=8cm,  clip]{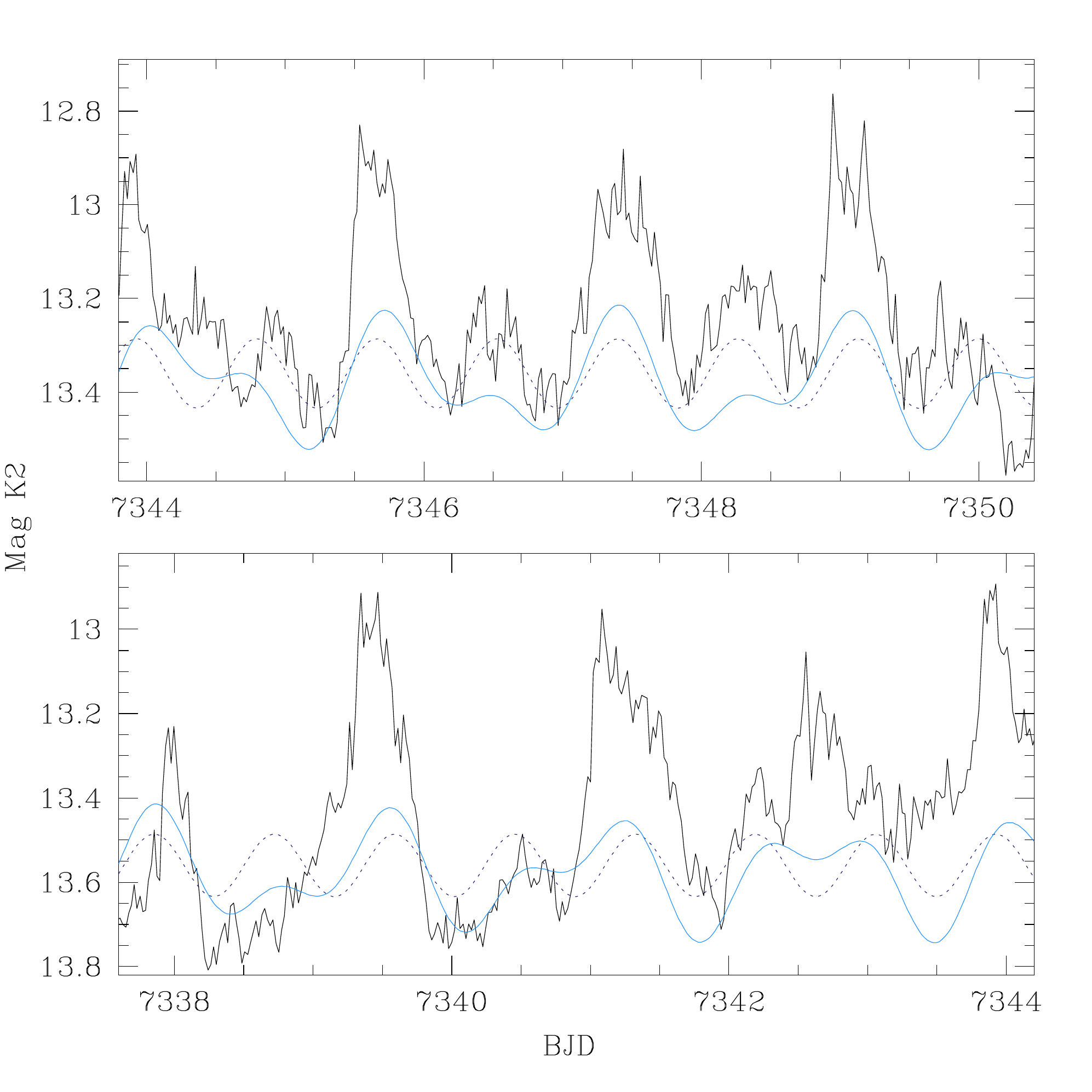}
  \caption{Portion of the light curve where the second, transient  f$_{0.63}$  frequency shows up stronger than the orbital period. 
  Apparently there are high-amplitude peaks in the light curve followed by a smaller peaks. Overall, this part of  the light curve is 
  better described by a double frequency curve than just orbital one, present throughout the data set. The source of the second frequency is unknown.   }
  \label{fig:2f460}
\end{figure}

The short cadence data are important to explore high-frequency variability. Hence, the low-frequency variability was removed 
from the data by subtracting the 
corresponding fit similar to the long cadence light curve, as described in the previous section.  However, no significant peaks 
other than those mentioned above are observed  in the short cadence data.  But, since the amplitude of the variability is changing 
with the overall brightness
of the system, we wanted to check for any possible transient frequencies in the light curve.   
Considering that the analyzed data is evenly sampled without any time gaps, there was no need to use an elaborate  wavelet analysis like one 
proposed by \citet{1996AJ....112.1709F}. Instead, we proceeded with a  method known as  the "sliding periodogram" described in detail by 
\citet[][and references therein]{2005MNRAS.362.1472N}. 
Thus, we simply divided the entire time series into 857 pieces of 10 day duration, with a 2\,hr step following the 
prescription\footnote{in the retrospect, there are no frequencies to explore below 2\,hr in the data.} between individual pieces.
We performed a period analysis on each 10-day-long piece and obtained 857 power spectra, which were stacked into a dynamical power density spectrum (DPDS).  
The DPDS at low frequencies is presented in Figure\,\ref{fig:wvlt1}. 
Any strictly periodic signal 
must appear as a persistent  feature at a constant frequency in the DPDS. As in the case with a simple periodogram,  we could not detect any significant lines. 
The worm-like features abundant in the  DPDS are random super-position of noise in a low-contrast image deprived of a strong signal, and are common in similar 
situations \citep[][]{2005MNRAS.362.1472N}. 
Several features, however, are  eye-catching.  The orbital period is seen throughout the entire dataset, but the signal is much stronger when the system 
is bright. In the central panel of Figure\,\ref{fig:wvlt1} depicting low frequencies, the dark blue vertical line indicates the orbital frequency. In the bottom 
panel, the power spectrum of the entire dataset is shown by the black line and a power spectrum obtained close to the peak brightness 
of the object by a violet line. That occasion is indicated by a horizontal violet line in the central panel.  There is a strong transient signal at 0.63 cycles/day 
in that part of the light curve, exceeding the orbital period. In the left panel the black line is the polynomial fit to the light curve, the violet is 
the amplitude of the signal at the orbital frequency, and the cyan is the amplitude of the f$_{0.63}$  frequency.  We have no explanation 
whether this periodicity is real and why it appears at the height of the brightness. Certainly the light curve at that stretch of time is better 
described by the presence of both frequencies as can be seen in Figure\,\ref{fig:2f460}. Here the violet dotted line is the Fourier component at the orbital period and 
the light blue line is the sum of both. During several nights the light curve is seen as having bimodal \gt{frequency}. 
A modulation due to two-pole accretion on the WD hardly can explain that  because it can modify the phase and the amplitude, but the period should be the same as orbital. 
We also have to beware that in optical observations of CVs,  \citet{1989IBVS.3383....1W} notes  random accretion processes that can 
produce periodic signals in short time segments.

\begin{figure}[!t]
\centering
 \includegraphics[width=8cm,  clip]{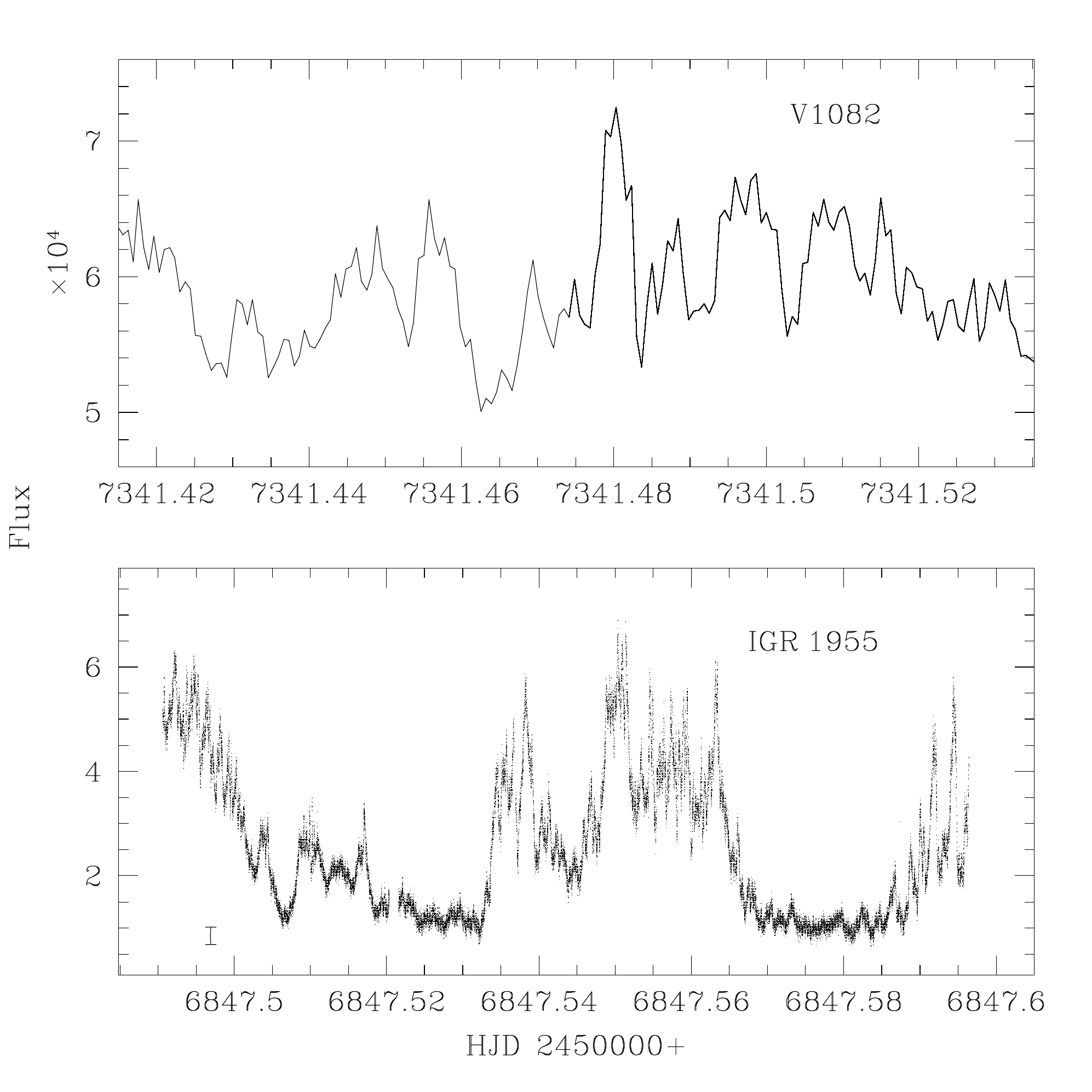}
  \caption{The light curve of \sgr\ in the active accretion mode is remarkable for its high-amplitude fast variability (top panel). That is not very unusual, though, 
  considering that the WD accretes the matter on its magnetic pole, similar to polars.   In the bottom panel a light curve of polar 
  IGR\,J19552+0044 \citep{2017A&A...608A..36T}  is presented 
  in a similar time and amplitude scale to demonstrate similarities.}
  \label{fig:sgrvigr}
\end{figure}

The speed and amplitude of the variability is not uncommon 
for magnetic WDs in polars. In the upper panel of Figure\,\ref{fig:sgrvigr} we plotted another arbitrary piece of the light curve of \sgr\  selected to 
match roughly  fast ground-based photometry of a known  polar IGR J19552+0044 \citep{2017A&A...608A..36T}, shown in the bottom panel. 
In fact, the photometry of  IGR\,J19552+0044 is faster and the  $K2$ light curve of \sgr\ looks like a smoothed version of the former. 

\begin{figure*}[t]
 \includegraphics[width=12cm, bb=50 40 580 780, angle=-90, clip]{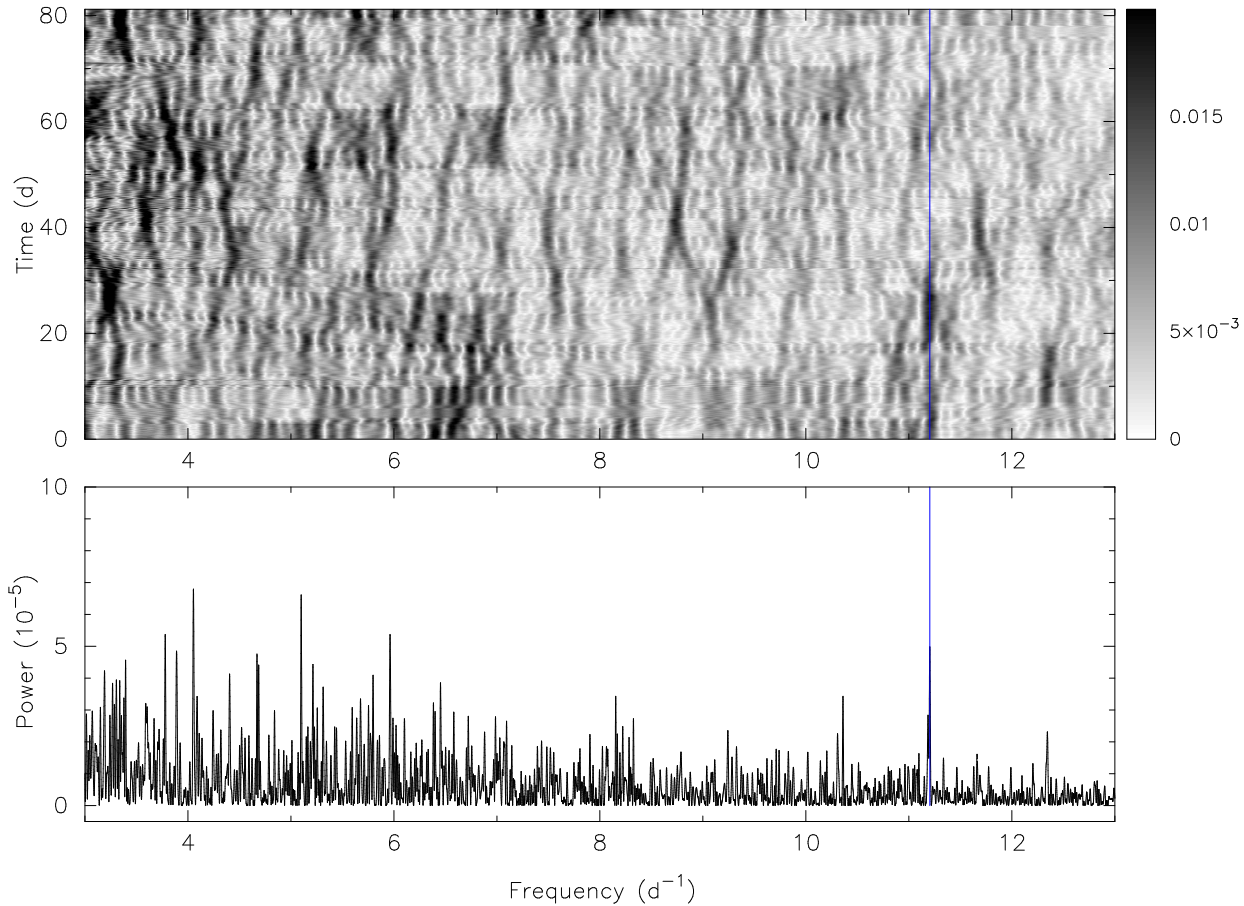}
  \caption{ Higher frequency portion of the wavelet. The strongest peak in this portion (two orders of magnitude smaller than the orbital frequency) is at 
  around 4 cycles/day. It is most likely an alias produced by the thruster firings  of the $K2$ satellite \citep{2016PASP..128g5002V}.  
  The other relatively strong and persistent peak is at  11.2 cycles/day. This one is remarkable for its possible counterpart in X-ray data
   \citep{2013MNRAS.435.2822B}. No other definitive frequencies are observed.
 }
  \label{fig:wvlt2}
\end{figure*}

Finally, the inspection of the wavelet in the higher frequencies (Figure\,\ref{fig:wvlt2})  shows just one artifact at f=11.2 cycles/day that is relatively strong and persistent. 
This signal is of interest as there was a transient periodic signal with a frequency of $2\pm0.14$\,h  detected in X-ray observations of V1082 Sgr \citep{2013MNRAS.435.2822B} 
that is close enough to suspect coincidence.
This timescale is intriguing as it is of the same order of magnitude as the estimated time it would take matter to travel from the donor star to the WD.  
We can claim a marginal detection of the same 
periodicity. It is the strongest in a time segment when the object is not at the maximum of brightness, but is active.

\section{Spotting the donor star}

\begin{figure*}[!t]
\centering
 \includegraphics[width=12cm,  bb=30 160 580 700, clip]{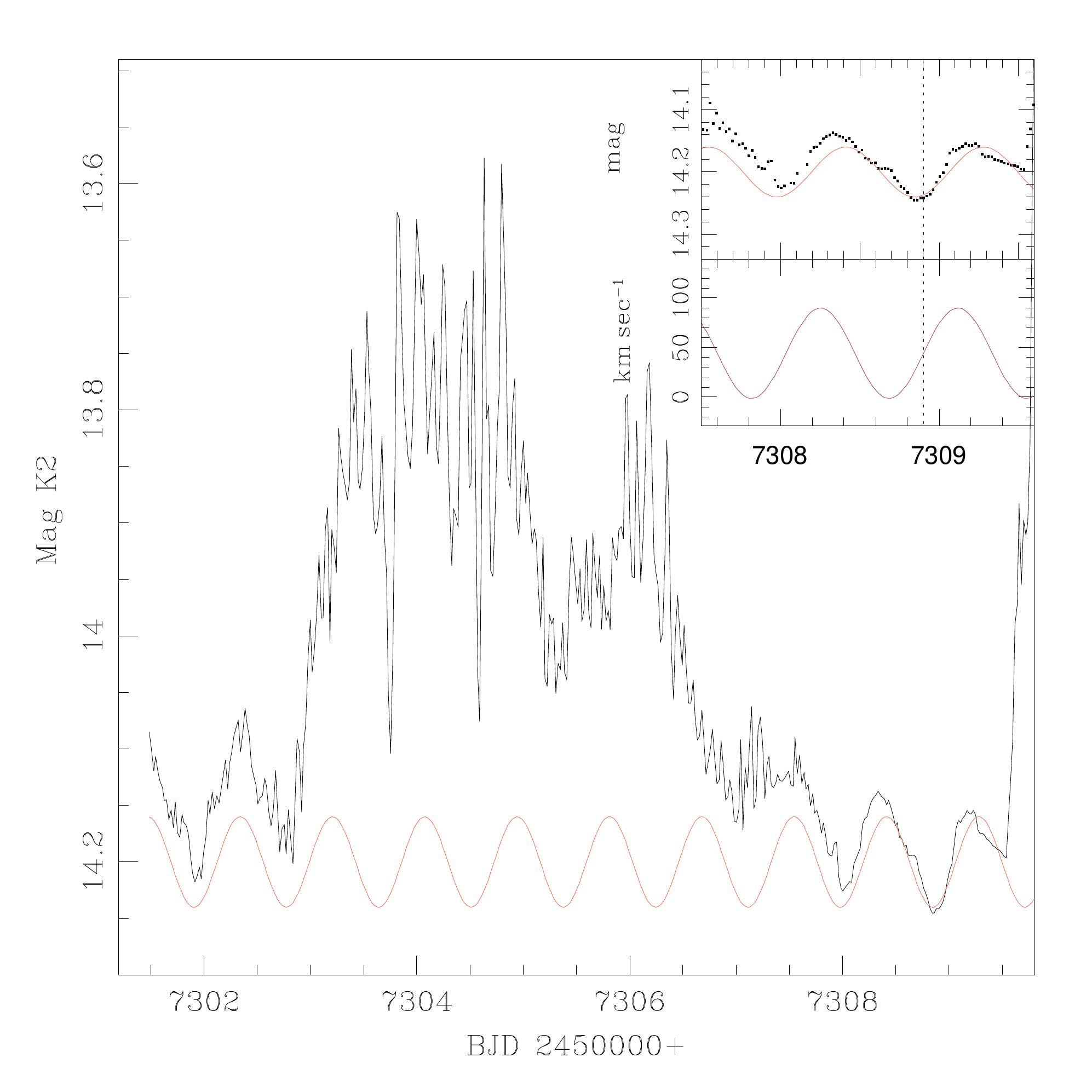}
  \caption{Another portion of the light curve of the short active accretion period flanked by two deep minima. The one at the right side starting around 
  BJD 7307.5 is particularly remarkable.  
  It is the lowest brightness observed in 80 days of monitoring and lasting 2.5 orbits ($\sim50$\,hr). The curve is smooth, without signs of accretion flickers, 
  and is clearly modulated with the orbital
  period. The orbital period is shown as a light red sinusoid fitted to that small portion of the light curve, presented in the inset of the figure in the upper panel.  
  In the bottom panel of the 
  inset a radial velocity curve from spectroscopic observations is plotted  (Paper\,II). The conjunction of stellar components is marked by a dotted vertical line.      }
  \label{fig:forbmin}
\end{figure*}

In the previous section we showed that the orbital period is omnipresent in the data, but often concealed  under other large amplitude variability. 
The most interesting part of the data is related to the minima in which the orbital modulation becomes  dominant.
There are three episodes related to the 29 day cyclical activity in the $K2$ light curve marked with red arrows
(see Figure\,\ref{fig:xlc}). The first one (near BJD 2457308) coinciding with the object reaching the faintest brightness level during  80 days, is of particular interest. 
Figure\,\ref{fig:forbmin} shows a portion of the $K2$ light curve 
when the deep minimum occurs.    As the brightness nears its  lowest levels, the flickering disappears completely, and the light curve is smooth and near-sinusoidal with a slight downward trend.  

A sinusoidal function with a fixed orbital frequency, but variable phase and amplitude, was fit to the small portion of the light curve around the minimum (BJD 7307.5 to 7309.5). The best fit model was estimated using a Monte Carlo Markov Chain (MCMC) implemented in python using EMCEE \citep{2013PASP..125..306F}. A total of 20 walkers were used with 25000 steps allowed, and the first 5000 steps were discarded as burn-in. Flat priors were assumed for the amplitude, phase, and offset. The best-fit model is shown in the top panel of the inset in Figure 8. In the inset's lower panel the RV curve corresponding to the donor star is plotted according to the precise RV measurements of recent high-resolution spectroscopic observations (Paper II). The ephemeris derived from the RV data predicts that minimum light when the light curve is dominated by the secondary star should be at orbital phase 0.25 (marked as dashed lines in the inset). However, the best-fit phase from the MCMC analysis was 0.3172$\pm$0.0001, which corresponds to a phase shift of 0.0672$\pm$0.0001, relative to the RV ephemeris. This error is dominated by the errors on the {\sl Kepler} data, and is likely too small given that the light curve is not perfectly sinusoidal. To account for this, we redid the MCMC analysis, but also allowed for a scaling factor on the error of the data with a Jeffreys prior (that is, $P(\sigma)\propto\frac{1}{\sigma}$. The resulting best-fit phase was 0.32$\pm$0.03 at the 3$\mathrm{\sigma}$ level, which corresponds to a phase shift of 0.07$\pm$0.03 relative to the RV ephemeris.

In the deep minimum, like the one achieved  in the selected portion of the light curve, the predominant source of light
has to be the donor star.    It can be safely assumed that the flickering is associated with the ongoing accretion, and 
that its disappearance  indicates  suspension of accretion-related radiation.  
\sgr\ was observed spectroscopically over extended periods of time, and
on several occasions it showed significant weakening of emission-line intensities  simultaneously with the fading of the brightness (continuum).
During short instances,  only a very narrow faint chromospheric H$_\alpha$ line is detected, indicating a total cessation  of accretion fueled ionization 
\citep{2016ApJ...819...75T}, while the continuum corresponds to a K-star  in the entire optical range. These observations lead to an important 
conclusion that the light from the donor star in this binary varies with the orbital period  with an amplitude of 0.08 mag amplitude.

\section{Estimates and discussion}

Detection of a single-humped, smooth, almost  sinusoidal wave  from the donor star in a compact binary means that the star's
rotation is synchronized  with the  orbital period. Even more importantly, it means  that the 
donor star is not ellipsoidaly deformed, i.e, does not fill its Roche lobe. In a Roche lobe overflowing long-period CV, in which the secondary star contributes 
significantly to the light, the variability is double-humped and usually uneven, like in V630 Cas  \citep[see Fig 5][]{2001MNRAS.326.1134O}. This confirms our
earlier interpretation of \sgr\ as a detached binary \citep{2017ASPC..509..489T}.

The variability then is a result  of a spot
covering  a large fraction of the stellar surface to produce the observed  amplitude of the light curve. It is typical of a fast rotating K-type dwarf in the form and the amplitude 
\citep[e.g. AB\,Dor][]{1992PASAu..10...33A}.
So, the natural assumption is  that the variability is caused by a a cool spot commonly observed in chromospherically active K stars. An alternative explanation could be
a possible heated/irradiated face (the side permanently facing the WD) of the  donor star. The expected phase shift in such a case 
between the RV zero 
point and the light curve extrema should be 0.25, instead of the observed 0.32 P$_{\mathrm {orb}}$ range. 
At this point it is difficult to say if this difference is significant enough 
to discard the latter possibility.

\begin{figure}[!t]
\centering
 \includegraphics[width=8cm,  clip]{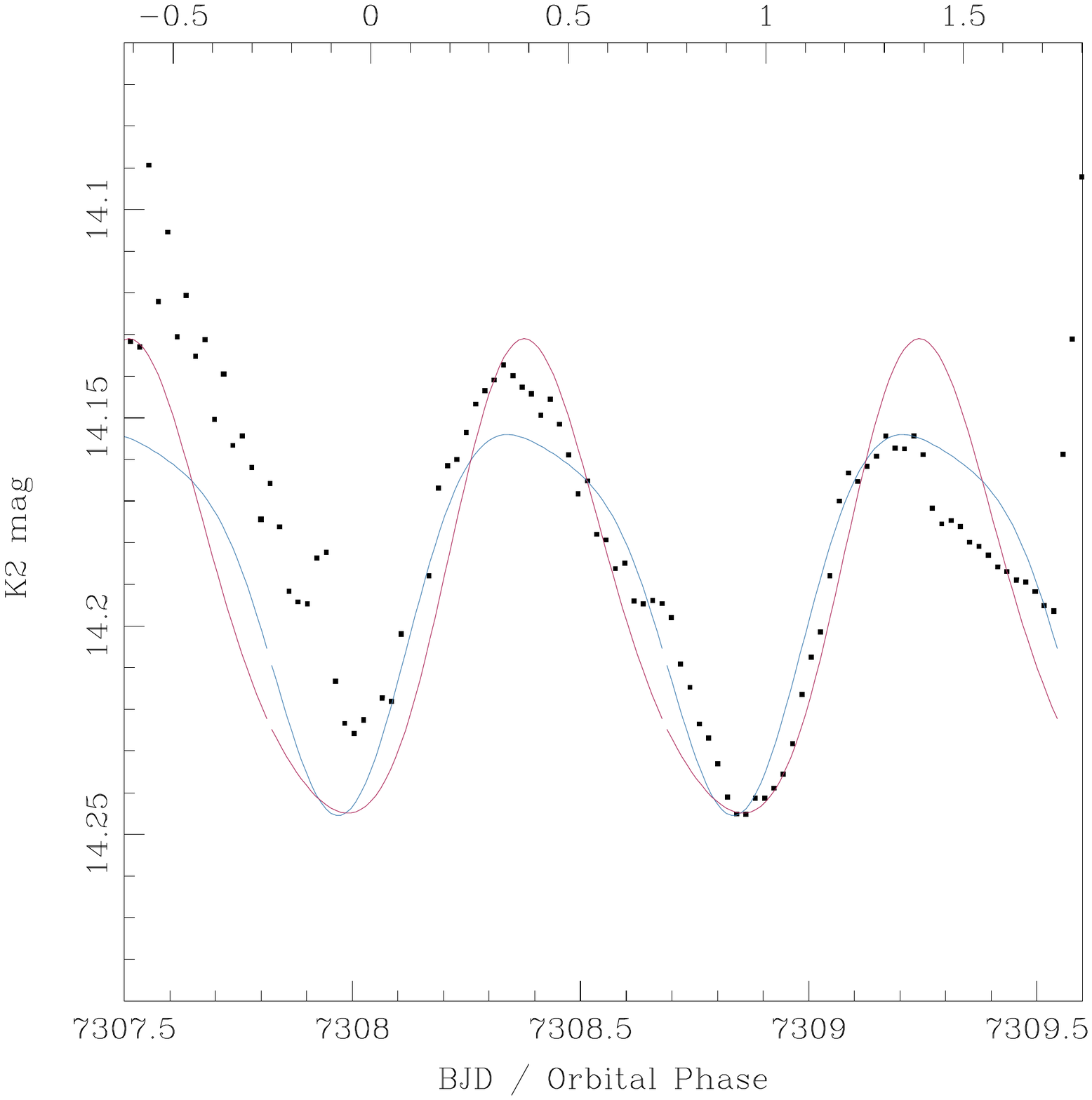}
  \caption{Light curve of the object at the deep minimum around BJD 7308.5. Black points are observational data and the 
  red curve is a binary model with hot spot on the face of the donor star (slightly shifted from line connecting the stars). 
  The blue line is a model with a cool spot on the surface of the donor star. The models are not fits to the actual observations,
  they are just made to qualitatively correspond to the shape of the light curve and amplitude. See the text for explanations.}
  \label{fig:sec}
\end{figure}

Nevertheless, we used a binary star modeling tool, Nightfall  \citep{2011ascl.soft06016W}, to
check for possible solutions.  Two and a half orbits of data with changing profile of the variability is not sufficient for a serious modeling attempt, hence we do not really fit the light curve.
Instead, in Figure\,\ref{fig:sec} we present the same fraction of the light curve, as in the inset of  Figure\,\ref{fig:forbmin},  by overplotting possible solutions offered by  Nightfall. 
Input parameters of the binary used to produce this light curve are listed in Table\,\ref{tab:binpar}. They were deduced  largely using the X-ray spectral 
fit \citep{2013MNRAS.435.2822B} for
the estimate of WD mass and the analysis of high-resolution spectroscopy in regard to the parameters of the donor star (see details in Paper\,II).

The probe confirms that employing a  fill factor $> 0.7$ immediately produces double humps in the light curve, even for such a low inclination of the orbital plane as $i=23$\grad.
The only viable solutions to reproduce the observed light curve involve a rather spherical star with spot(s).   
  
The red symmetric curve in Figure\,\ref{fig:sec} is produced by adding a heated spot in the front of the star. It resembles  the first full orbit at the 
deep low state between BJD 7308 -- 7309. 
However, there is a "knee" on the receding wing of the light curve, which is  much stronger in the second orbit  between BJD 7309 -7309.6 after which the system suddenly brightens up. 
The blue model (the blue curve in Figure\,\ref{fig:sec}), which was calculated by adding a cool spot at the back of the K-star, seems to work better to explain the asymmetric wing. 
The problem is not only with the symmetry. In order to place the minimum light 0.06 orbital phases prior to the conjunction (corresponding to the orbital phase 0.0), 
the heated spot should be displaced from the center of the line connecting 
 stellar components by  as much as $35^{\circ}$ in longitude. To make it wide enough, the heated spot has a $\sim50^{\circ}$ deg radius, which is not really realistic. 
 In contrast, the cool spot should not necessarily be aligned with anything and does not have to be round or symmetrical. It also can be variable in size and intensity. 
 
 We think that a cool spot is a more 
 reasonable solution to explain the observed variability of the K-star. The parameters of the cool spot in the presented model assume a spot that is 370\,K cooler than 
 the effective temperature of the star.   The spots in such active 
 stars can be really long lived, but they  evolve and patterns change. Usually, the spots are neither single nor round. 
 So the stellar surface cannot be  accurately modeled  based on the short duration of deep minima.  Neither can  a combination of a hot spot with a cool one be excluded. 
 We only attempted  to show that  a spot assumption  is realistic in terms of its  size and temperature difference.
  
 \begin{table}[h]
\caption{Binary parameters used in the Nightfall}
\begin{center}
\begin{tabular}{lccc}
\hline
\hline
Inclination &     23.0 &  &  (degree) \\
Mass Ratio &       0.88 & &   \\
Period &           0.867 & &  (days) \\
Total Mass  &       1.37 & &  (solar mass) \\
Separation &        4.25 & &  (solar radius) \\
\hline
\hline
         &         Donor  & WD  &  \\
\hline
Amp. RV &    45.35  &    51.5  &  (km/s$^{-1}$) \\
Mass &             0.73  &    0.64  &  (solar mass) \\
Mean Radius &      1.1   &   0.015 &  (solar radius) \\
Fill Factor   &      0.67  &    0.01  &   \\
Temperature &    4930.0 &   30000.0 &  (Kelvin) \\
Asynchronicity &  1.0   &   1.0 &   \\
Albedo   &           0.5    &  1.0   &   \\
Grav. Dark. &      0.107    &  0.25  &   \\
\hline
\hline
Spot on the &  donor  &star   & \\
 Long. & Lat.  &  Radius & DimFactor \\
 \hline
 145.0   &  0.0  &  50.0   &  0.925 \\
 \hline
\end{tabular}
\end{center}
\label{tab:binpar}
\end{table}%

\section{Conclusions}

We explored an unprecedented 80-day-long uninterrupted light curve of the enigmatic binary \sgr\ with high time resolution
obtained by the $K2$ mission. This is  impossible  using ground-based telescopes, with 
the  long 0.867\,d orbital period of the binary and $\sim29$\,days cyclical variability.
We identified the orbital variability in the light curve known previously from spectroscopy. We also observe cyclical, 
quasiperiodic minima in the light curve in the $K2$ data, as well as in the follow-up ground based photometry. The minima usually are brief,
lasting two or three orbital periods, followed by a sudden and strong increase of the brightness. When the system is active it
exhibits fast, large amplitude variability resembling  ordinary magnetic CVs or polars.  Some transient periods are seen 
in the active phase, for which there are no  ready explanations. The $\sim2$\,hr transient period that has been observed in X-rays 
is marginally detected in the optical light curve. 

The only persistent, and strictly periodic signal is the orbital one, revealed in both active accretion and accretion-shutdown states. 
In the active state the signal at the orbital frequency is strong, indicating  modulation of the accretion process similar to polars. 
The period is coherent throughout the entire 80 day dataset. 

In the low brightness state the orbital period persists and is clearly visible. 
Spectroscopic and photometric observations  and the spectral energy 
distribution  show  that during this interval nearly all of the light comes from the K\,2 star. 
Detection of a smooth orbital variability during the deep minimum  confirms the absence of accretion, as the latter is usually 
accompanied with fast spikes in the light curve. The form of the light curve suggests that the K-star  is not ellipsoidally shaped.  
The duration of this nonaccreting state is not long and has not been observed repetitively with sufficient time resolution  in order 
to provide an observational base for  serious modeling. However, we demonstrate qualitatively that it can simply be  a result of a large cool spot on the
surface of a chromospheric K-star. 

The orbital variability  could also be produced by a hot spot formed as a result of irradiation.   
The reality can be more complex with both hot and cool spots present.

Regardless of the origin of spot(s), the light from the K-star modulated with the orbital period means that the K-star rotates 
synchronously with the binary system.  The  form of the variability   indicates an absence of deformation
associated with the star filling its Roche lobe. These are the two the most important results from our analysis of the $K2$ observations.  
Paper\,II pairs these conclusions  with the mass of the WD determined from the X-ray
observations  and the results of the high-resolution spectroscopy  to further the  interpretation of this interesting system.

\acknowledgments
G.T.  acknowledges PAPIIT grants IN108316 and CONACyT grant 166376. P.S. acknowledges NSF grant AST-1514737.  M.R.K. is funded
through a Newton International Fellowship provided by the Royal Society.

\facilities{{\sl Kepler-K2}, OAN SPM (RATIR)}
\software{IRAF, Everest, Nightfall, Period04}

\end{document}